# Constructor Theory of Thermodynamics

Chiara Marletto[1]

Materials Department, University of Oxford

May 2017

Current formulations of the laws of thermodynamics are valid only at 'macroscopic scales', which are never properly defined. Here I propose a new scale-independent formulation of the zeroth, first and second laws, improving upon the *axiomatic approach to thermodynamics* (Carathéodory, 1909; Lieb & Yngvason, 1999), via the principles of the recently proposed *constructor theory*. Specifically, I provide a scale-independent, more general formulation of 'adiabatic accessibility'; this in turn provides a scale-independent distinction between work and heat and it reveals an unexpected connection between information theory and the *first* law of thermodynamics (not just the second).. It also achieves the long-sought unification of the axiomatic approach with Kelvin's.

1. Introduction

An insidious gulf separates existing formulations of thermodynamics from other fundamental physical theories. They are *scale-dependent* – i.e., they hold only at a certain 'scale', or level of 'coarse-graining', none of which are ever exactly specified. So existing thermodynamics provides unambiguous predictions about 'macroscopic' systems such as Victorian heat engines, but it is controversial how it applies to 'microscopic' ones, such as individual quantum systems.

---

[1] Address for correspondence: chiara.marletto@gmail.com

Here I propose a *scale-independent* formulation of the zeroth, first and second laws of thermodynamics – i.e., one that does not rely on approximations, such as 'mean values on ensembles', 'coarse-graining procedures', 'thermodynamic equilibrium', or 'temperature'. This new approach uses the principles and tools of the recently proposed *constructor theory* [1], especially the *constructor theory of information* [2].

I shall present the laws following the tradition of foundational studies in thermodynamics initiated by Carathéodory in 1909 [3] and improved upon by Lieb&Yngvason's [4] – the so-called *axiomatic thermodynamics.* This is because it can be more easily generalised in constructor theory, in that as I shall explain it is based on requiring certain transformations to be impossible. Specifically, the second law requires certain states of a physical system to be 'adiabatically inaccessible' from one another. One of the key innovations of this work is the constructor-theoretic definition of adiabatic accessibility, which generalises Lieb&Yngvason's [4] and makes it scale-independent, thus also providing a scale-independent distinction between work and heat. I also provide a new, exact link between thermodynamics and (constructor-theoretic) information theory, not only via the second law, as one would expect [5, 6], but also via the *first*. As a result, axiomatic thermodynamics and Kelvin's more traditional approach are unified.

## 1.1. The role of constructor theory

In constructor theory all physical laws are expressed exclusively via statements about which physical transformations, or '*tasks*' (see section 2), are *possible*, which are *impossible*, and why. This mode of explanation sharply differs from *the traditional conception of fundamental physics*, under which physical laws are to be expressed by stating what *must happen*, given boundary conditions in spacetime that sufficiently fix the state.

Constructor theory is not just a framework (such as resource theory, [7], or category theory [8]) for reformulating existing theories: it also has new laws of its own, *supplementing* existing theories. These are new *principles* – laws about laws, intended to underlie other physical theories (e.g. elementary particles' dynamical laws, etc.), called *subsidiary theories* in this context. These principles express subsidiary theories' regularities, including *new* regularities that the traditional conception cannot adequately capture.

Specifically, I shall appeal to the principles of the *constructor theory of information* [2]. They express the regularities that are implicitly required by information theories (e.g. Shannon's), via scale-independent statements about possible and impossible tasks, thus giving full physical meaning to the hitherto fuzzily defined notion of information. The principles are crucial (see section 5) to express a scale-independent connection between thermodynamics and constructor-theoretic information theory.

Central to this paper is the difference between *a task* being *possible* and *a process being permitted* by dynamical laws. The latter means that the process occurs spontaneously (i.e., when the physical system has no interactions with the surroundings) under the dynamical laws, given certain boundary conditions. In contrast, a task is 'possible' if the laws of physics allow for arbitrarily accurate approximations to a *constructor* for the transformation that the task represents. A constructor (section 2) is an object that, if presented with one of the task's designated inputs, produces (one of) the corresponding outputs, and *retains the ability to do this again*. Thus it must operate 'in a cycle'. The concept of a constructor is extremely general; for example, actual computers, heat engines and chemical catalysts are approximately-realised constructors. In reality no perfect constructor ever occurs, because of errors and deterioration; but whenever a task is possible a constructor for that task can be approximated to arbitrarily high accuracy. Under constructor theory (despite its name!) laws are expressed referring exclusively to the possibility or impossibility of tasks, *not* to constructors.

Accordingly, the constructor-theoretic second law will be stated as the *impossibility* of certain tasks, as in Kelvin's, Clausius's and Lieb&Yngvason's formulations. This is in sharp contrast with the statistical-mechanical approach, where the second law concerns *spontaneous processes* on isolated, confined systems.

## 1.2. The problem

There are several not-quite-equivalent formulations of thermodynamics [9, 10]. All of them have in common the fact that their laws are scale-dependent – i.e., their domain of applicability is limited to a certain scale, which is never properly defined. Since my approach improves particularly on the axiomatic formulation of thermodynamics, I shall refer to it (in its most recent formulation, by Lieb and Yngvason [4]) to illustrate the problem of scale-dependence.

In thermodynamics one usually distinguishes between the system and its surroundings. They can be in a number of thermodynamic states – corresponding to equilibrium states (see below). The primitive notion of the axiomatic formulation is that of an *adiabatic enclosure*:

> An adiabatic enclosure is one that allows for a physical system inside it to be perturbed by mechanical means [see below] only.

where it is customary to assume that 'mechanical means' correspond to any change in the surroundings equivalent to the displacement of a weight in a gravitational field. The *first law of thermodynamics* then states that all ways of 'doing work' on adiabatically enclosed systems (i.e. those inside an adiabatic enclosure) are equivalent [9,10] in the sense that:

> The work required to change an adiabatically enclosed system from one specified state to another specified state is the same however the work is done.

Consequently, one introduces internal energy as an additive function of state, obeying an overall conservation law. Heat is then defined in terms of work (see, e.g., [12]): the 'heat' absorbed by the system while it is driven from the state $x$ to the state $y$ is defined as the difference $δQ$ between the work $ΔU$ required to drive it from the state $x$ to the state $y$ when adiabatically enclosed, and the work $δW$ required to drive it between the same two states without the adiabatic enclosure. The classic expression of the first law is then:

$$\Delta U = \delta W + \delta Q \qquad (1)$$

The second law is rooted in the notion of *adiabatic accessibility* [4] of states $x$ and $y$, and can be expressed as follows:

> The state $y$ is adiabatically accessible from the state $x$ if the physical transformation $\{x \to y\}$ can be brought about by a device capable of operating in a cycle [a constructor], with the sole side-effect (on the surroundings) being the displacement of a weight in a gravitational field.

The second law is then formulated as [3]:

> In any neighbourhood of any point $x$ there exists a point $y$ such that $y$ is not adiabatically accessible from $x$.

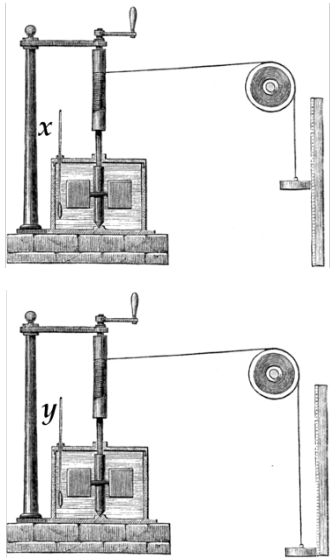

**Figure 1**: The state $y$, at a higher temperature, is adiabatically accessible from $x$, at a lower temperature, but not vice-versa. (Adapted from: *Harper's New Monthly Magazine*, No. 231, August, 1869).

An example of such $x$ and $y$ is given by Joule's experiment (see **Figure 1**) to measure 'the mechanical equivalent of heat'. $x$ and $y$ are two thermodynamic states of some volume of water – labelled by their different temperatures $x$ and $y$. Under known laws of physics, if $y > x$, $y$ is adiabatically accessible from $x$, but not vice versa. Here, the constructor to bring the fluid from temperature $x$ to temperature $y$ adiabatically consists of the stirrer and the pulley (as they undergo no net change and can work in a cycle), and the 'weight' includes the string.

The form of these laws is scale-dependent for two main reasons.

First, they rely on the existence of 'equilibrium states' (or equivalent) for the system in question. Specifically, equilibrium states are postulated via two routes. One is the so-called *"minus-first" law* – [16]:

> An isolated system in an arbitrary initial state within a finite fixed volume will spontaneously attain a unique state of equilibrium.

This is a tacit assumption of all formulations of thermodynamics, including the axiomatic one [16]. The other is the *zeroth law* (see e.g. [9]), which is used to define temperature by requiring transitivity of the thermodynamic equilibrium relation between any two physical systems. However, the equilibrium condition is only ever satisfied at a certain scale. For instance, equilibrium states can be defined as those "which, once attained, remain constant in time thereafter until the external conditions are unchanged" [16]. Such states are never exactly attained, because of fluctuations (both in the classical [18] and quantum [19] domains). Thus, by relying on the existence of equilibrium states via the above two laws, the first and second laws of thermodynamics do not hold at all scales: no confined physical system admits equilibrium states as defined above, unless when considered over a suitably defined time-scale, which can only be approximately defined.

In addition, the notions of 'mechanical means' and 'adiabatic accessibility' are themselves scale-dependent: they appeal to macroscopic objects such as weights in a gravitational field or equivalent. Hence, the laws of thermodynamics in their current forms are not directly applicable in general situations – such as in information-processing nanoscale devices (see, e.g., [14, 15]), where side-effects of transformations are not clearly related to a weight or other intuitively defined mechanical means. There are cases where for instance it is impossible to couple a weight to the system in order to perform a transformation; but nevertheless that transformation can be performed adiabatically. For example, the task of setting a quantum particle initially prepared in an energy eigenstate $|E_0\rangle$ to a different eigenstate $|E_1\rangle$ can be performed adiabatically, e.g. by implementing a unitary transformation $|E_0\rangle|E_1\rangle \to |E_1\rangle|E_0\rangle$ on two such quantum particles, but not by direct coupling to 'mechanical means'. Thus, to let the first and second laws to cover such cases one needs a more general, scale-independent notion of 'adiabatic accessibility' and of 'mechanical means'. My paper provides both.

Deriving thermodynamics from statistical mechanics [20] does not solve the scale-dependence problem. Statistical mechanics' aim is to reconcile the second law with the traditional conception's approach (where everything is expressed in terms of predictions and laws of motion). Thus, the second law in this context is cast as requiring there to be irreversibility in *some spontaneous evolution* of *confined, isolated systems* [21]. Such irreversibility is usually expressed in terms of *entropy* being a globally non-decreasing function [13]. However, deriving or even reconciling such statements with the microscopic time-reversal symmetric dynamical laws is notoriously problematic [21], because of Poincaré recurrence in confined systems [22]. There are a number of *models* where time-reversal asymmetric, second-law like behaviour arises from microscopic time-reversal symmetric laws. Such models, however, produce scale-dependent predictions, for instance because they are falsified by the existence of fluctuations. In addition, some of them adopt approximations involving ensembles or coarse-graining procedures, or statistical (probabilistic) assumptions about the process of equilibration – e.g. its "probably" leading to the "most probable" configuration – defined with respect to a natural measure on phase space; or about some specially selected, ad hoc, initial conditions. Resorting to such approximation schemes only means, yet again, that the laws hold at some non-specified scale. Hence, the statistical-mechanics path to a scale-independent foundation for thermodynamics is no less problematic.

Now, the above problems are generally regarded as unsolvable: thermodynamic laws are expected to hold only at macroscopic scales only. This capitulation in the face of foundational problems generates a number of troubling open issues. For instance, updated versions of Maxwell's demon (e.g. Szilard's engine [23]), purporting to violate the second law of thermodynamics, are hard to exorcise. In such models, the working medium of the alleged perpetual motion machine of the second kind is constituted by a *single* particle. Since the second law is only known in scale-dependent forms, it is difficult to pin down what exactly it is that exorcises the demon: it is controversial what exactly the second law forbids in that context [6, 24]. Similar problems arise in the new field of quantum thermodynamics [14, 15], which investigates the implications of thermodynamics for quantum systems such as atomic-scale 'heat-engines'. In addition, the known connection between existing information theory (classical and quantum) and thermodynamics [5, 6] – together with the isolated case of the entropy of an individual black hole [25] – strongly suggest that there is indeed a scale-independent formulation of thermodynamic laws.

### 1.3. The logic of the solution

In this paper I propose a scale-independent formulation of the zeroth, first and second laws of thermodynamics.[2] In particular, they do not rely on 'coarse-graning', 'equilibrium' or 'thermodynamic limit'; and they do not rely on macroscopic objects such as weights in a gravitational field. The consistency with existing laws will be explained by means of examples, as the laws are presented in the paper.

The logic goes as follows. First, the existence of equilibrium states does not need to be postulated (see section 2.1) and the constructor-theoretic zeroth law (see section 6) will be expressed as a scale-independent statement about certain tasks being possible (without relying on there being equilibrium states, nor temperature). Note that such a statement does not require the *existence* of ideal constructors (see section 2.1); rather, it refers exclusively to physical laws allowing a *sequence* of arbitrarily accurate approximations to ideal constructors.

---

[2] The problem of scale-dependence affects the third law of thermodynamics too – for, in traditional thermodynamics, that law is formulated either in terms of temperature, or in terms of the behaviour of the entropy function near temperature zero [10]. I shall leave a scale-independent formulation of the third law for future work – it is expected to follow from the other laws of thermodynamics and the principles of constructor theory.

In addition, I shall formulate scale-independent expressions of 'adiabatic accessibility' and of 'mechanical means' (section 5). Specifically, I define a class of physical systems – *work media*, generalising the notion of 'mechanical means' — by stating what tasks must be possible on them. This class would include idealised weights, springs and flywheels; but also, e.g., certain quantum particles in certain quantum states.

I then define 'adiabatic accessibility' in terms of work media. This definition generalises existing ones to any subsidiary theory obeying constructor theory's principles, e.g. even post-quantum theories, and it is scale-independent.

A scale-independent formulation of the first and second laws follows; this includes a definition of another class of substrates – *heat media* – which would include systems prepared e.g. in a thermal state – again by stating in constructor-theoretic terms what tasks must be possible on them. The well-known formula (1) shall be recovered as the ending point of our construction (see section 7), thus making contact with the existing formulations of thermodynamics. The key difference is that 'heat' and 'work' will be defined in a scale-independent way. Thus, the laws of thermodynamics formulated in this paper are applicable to nano-scale devices too – although the application of the laws to the particular case of quantum theory goes beyond the scope of this paper. Just like for classical thermodynamics, the laws are formulated in the form of principles, not derived from any specific dynamical law.

As I mentioned above, the laws I propose are not confined to equilibrium thermodynamics, nor do they rely on any notion of temperature. However, as I shall explain, heat media include systems prepared in a thermal state. The connection with equilibrium thermodynamics (which will not be explored in this work but is an interesting future application of it) may be established via the constructor-theoretic version of the *zeroth law* (section 6) – which as I mentioned is not about temperature, but about the possibility of certain tasks.

As explained in [11, 13], Kelvin's statement of the second law sets a definite direction to the 'irreversibility' of the second law, stating that it is impossible to convert heat completely into work (without any other side-effects), but it is possible to do the reverse; axiomatic-type statements are weaker, in that they only imply the irreversibility of some transformation, without specifying which direction is forbidden. In this work the two approaches are unified, and heat and work given scale-independent constructor-theoretic characterisations.

Expressing the laws in terms of possible or impossible tasks, (not about spontaneous processes happening) has the significant consequence that the statement of the second law requiring that a certain task (say, $\{x \to y\}$) be *possible*, but the task representing the inverse transformation, $\{y \to x\}$, be impossible, does not clash with time-reversal symmetric dynamical laws. For (see section 2.1) a task being possible, unlike a process happening, requires a constructor for the task; and the time-reverse of a process including a constructor for $\{x \to y\}$ need not include a constructor for $\{y \to x\}$.

Another consequence is that one can accommodate *counterfactual properties* of physical systems – about what can, or cannot, be done to them. This is the key to the scale-independence of the new notion of adiabatic accessibility: 'work media', on which the definition relies, are defined by their counterfactual properties.

In constructor theory entropy is not as central as in axiomatic thermodynamics. The second law's physical content is *not* that it forbids tasks that decrease (or increase) the entropy of a system, but that it forbids certain tasks to be performed adiabatically (as defined in section 5), while requiring the inverse physical transformation to be performable adiabatically. Entropy enters the picture *after* the second law, as a quantitative classification of tasks, so that tasks that change entropy by the same amount belong to the same class; but the physical content of the second law resides in the definition of adiabatic accessibility.

Finally, in this paper the *first* law of constructor-theoretic thermodynamics is connected to (constructor) information theory. This is a novel development, which has interesting implications for quantum information and quantum thermodynamics: in all previous treatments it is only the second law that is so connected. Moreover, information-based concepts such as distinguishability provide scale-independent physical foundations for the notions of heat and work, and for distinguishing between them. This is because in the constructor information theory (section 3), the fuzzily-defined traditional notion of information [27] is replaced by exact ones. In particular, the constructor-theoretic notion of information does not rely on any subjective, agent-based, or probabilistic/statistical statements about reality.

## 2. Constructor Theory

In traditional thermodynamics the system can be acted upon by a *thermodynamic cycle* – a sequence of changes in the surroundings that start and end in the same thermodynamic state. In constructor theory, physical systems are replaced by *substrates* – i.e., physical systems some of whose properties can be changed by a physical transformation brought about by a constructor. A *constructor* is in turn a substrate that undergoes no net change in its ability to do this – generalising a thermodynamic cycle. The primitive elements in constructor theory are *tasks* (as defined below), and statements about their being possible/impossible. The general descriptors of a substrates can be defined as follows:

**Attributes and variables.** For any substrate, subsidiary theories must provide its *states*, *attributes* and *variables*. States correspond to what in the traditional conception one would call '*microscopic states*' – so they should not be confused with thermodynamic states. For instance, in quantum theory a state is a particular density operator; in classical physics a point in phase space, etc.

An *attribute* of a substrate is formally defined as a *set of all the states* in which the substrate has a particular property; some attributes generalise thermodynamic states, as I shall explain. For example, a die on a table is a substrate. Its upturned face is a substrate that can have *six attributes* $n : n \in \{1,2,...6\}$, each one consisting of a vast number of states representing, say, the microscopic configuration of the die's atoms. One can construct other attributes by set-wise union or intersection. For instance, the attribute '*odd*' of the upturned face, denoted by ***odd***, is the union of all the odd-numbered attributes: ***odd* = *1* ∪ *3* ∪ *5***. Similarly for the attribute 'even': ***even* = *2* ∪ *4* ∪ *6***. Attributes generalise and make exact the notion of 'macrostates', or 'thermodynamic states'; crucially, attributes are not the result of any approximation (e.g. coarse-graining). An *intrinsic* attribute is one that can be specified without referring to any other specific system. For example, 'showing the same number' is an intrinsic attribute of a pair of dice, but 'showing the same number as the other one in the pair' is not an intrinsic attribute of either of them.

The notion of a thermodynamic property is generalised by that of a physical *variable*. The latter is defined in a slightly unfamiliar way as any *set of disjoint attributes* of the same substrate, each labelled by a value that the variable can take. Whenever a substrate is in a state in an attribute $x \in X$, where *X* is a variable, we say that ***X*** is *sharp* (on that system), with the *value x* – where the *x* are members of the set *X* of labels of the attributes in ***X***[3]. In quantum theory, variables include not only all quantum observables, e.g. angular momentum, but many other constructs, such as any set $\{***x***,***y***\}$ where the attributes ***x*** and ***y*** each contain a single state $|x\rangle$ and $|y\rangle$ respectively, not necessarily orthogonal. For example, a variable might be a set of attributes each corresponding to a quantum system being in a certain thermal state. Each attribute is then labelled by a value of temperature.

As a shorthand, "***X*** is sharp in ***a***" shall mean that the attribute ***a*** is a subset of some attribute in the variable ***X***. In the case of the die, 'parity' is the variable ***P*** = {***even***, ***odd***}. So, when the die's upturned face is, say, in the attribute ***6***, we say that "***P*** is sharp with value *even*". Also, we say that ***P*** is sharp in the attribute ***6***, with value *even* – which means that ***6*** ⊆ ***even***. In quantum theory, the *z*-component-of-spin variable of a spin-½ particle is the set of two attributes: that of the z-component of the spin being ½ and -½. That variable is sharp when the qubit is in a pure state with spin ½ or -½ in the z-direction, and is non-sharp otherwise.

---

[3] In this paper, I use this notation: Small Greek letters denote states; ***small italic boldface*** denotes attributes; ***CAPITAL ITALIC BOLDFACE*** denotes variables; *small italic* denotes labels; *CAPITAL ITALIC* denotes sets of labels; **CAPITAL BOLDFACE** denotes physical systems; and capital letters with arrow above (e.g. $\vec{C}$) denote constructors.

**Tasks**. A *task* is the *abstract specification* of a set of *physical transformations* on a substrate. It is expressed as a *set of ordered pairs of input/output attributes* $x_i \to y_i$ of the substrates. I shall represent it as:

$$\mathfrak{A} = \{x_1 \to y_1,\ x_2 \to y_2, ...\}.$$

The $\text{In}(\mathfrak{A}) \doteq \{x_i\}$ are the legitimate *input attributes*, the $\text{Out}(\mathfrak{A}) \doteq \{y_i\}$ are the *output attributes*. The *transpose* of a task $\mathfrak{A}$, denoted by $\mathfrak{A}^\sim$, is such that $\text{In}(\mathfrak{A}^\sim) = \text{Out}(\mathfrak{A})$ and $\text{Out}(\mathfrak{A}^\sim) = \text{In}(\mathfrak{A})$. A task where $\text{In}(\mathfrak{A}) = V = \text{Out}(\mathfrak{A})$ for some variable $V$ will be referred to as 'a task $\mathfrak{A}$ *over* $V$'. A task is fundamentally different from a thermodynamic process, because the latter is a particular trajectory the thermodynamic phase space that is permitted by the dynamical laws; whilst in constructor theory a task might represent an impossible transformation (see below), for which there is no such path.

A **constructor** for the task $\mathfrak{A}$ is defined as a physical system that would cause $\mathfrak{A}$ to occur on the substrates and *would remain unchanged in its ability to cause that again*. Schematically:

$$\text{Input attribute of substrates} \xrightarrow{\text{Constructor}} \text{Output attribute of substrates}$$

where constructor and substrates jointly are isolated. This may entail the constructor's changing its microscopic state: what is required is that it remain in the *attribute* of being capable of bringing the task about when given the legitimate input states.

A constructor is *capable of performing* $\mathfrak{A}$ if, whenever presented with the substrates (where it and they are *in isolation*) with a legitimate input attribute of $\mathfrak{A}$ (i.e., in *any* state in that attribute) it delivers them in *some* state in one of the corresponding output attributes, regardless of how it acts on the substrate when it is presented in any other attribute. For instance, a task on the die substrate is $\{even \to odd\}$; and a constructor for it is a device that must produce *some* of the die's attributes contained in ***odd*** whenever presented when *any* of the states in ***even***, and retain the property of doing that again. In the case of the task $\{6 \rightarrow odd\}$ it is enough that a constructor for it delivers *some* state in the attribute ***odd*** – by switching ***6*** with, say, ***1***.

A task $\mathtt{T}$ is *impossible* (denoted as $\mathtt{T}^x$) if there is a law of physics that forbids its being carried out with arbitrary accuracy and reliability by a constructor. Otherwise, $\mathtt{T}$ is *possible*, (denoted by $\mathtt{T}^{\checkmark}$). This means that a constructor capable of performing $\mathtt{T}$ can be physically realised with arbitrary accuracy and reliability (short of perfection). As I said, heat engines, catalysts and computers are familiar examples of *approximations* to constructors. So, '$\mathtt{T}$ is possible' means that it can be brought about with arbitrary accuracy, but *it does not imply that it will happen*, since it does not imply that a constructor for it will ever be built and presented with the right substrate. Conversely, a prediction that $\mathtt{T}$ will happen with some probability would not imply $\mathtt{T}$'s possibility: that 'rolling a seven' sometimes happens when shooting dice does not imply that the task 'roll a seven under the rules of that game' can be performed with arbitrarily high accuracy.

Constructor theory's *fundamental principle* is that

I. All (other) laws of physics are expressible solely in terms of statements about which tasks are possible, which are impossible, and why.

Hence principle I requires subsidiary theories to have two crucial properties: (i) They must define a topology over the set of physical processes they apply to, which gives a meaning to a sequence of approximate constructions *converging* to an exact performance of $\mathtt{T}$; (ii) They must be *non-probabilistic* – since they must be expressed exclusively as statements about possible/impossible tasks. The latter point may seem to make the task of expressing the laws of thermodynamics particularly hard, but that is only an artefact of the traditional conception (which tries to cast the second law as a model to provide predictions of what will happen to a system evolving spontaneously) that makes probabilities appear to be central to the second law. In fact none of the laws, in the constructor-theoretic formulation, use probabilistic statements.[4]

**Principle of Locality.** A pair of substrates $\mathbf{S}_1$ and $\mathbf{S}_2$ may be regarded as a single substrate $\mathbf{S}_1 \oplus \mathbf{S}_2$. Constructor theory requires all subsidiary theories to provide the following support for the concept of such a combination. First, $\mathbf{S}_1 \oplus \mathbf{S}_2$ is indeed a substrate. Second, if subsidiary theories designate any task as possible which has $\mathbf{S}_1 \oplus \mathbf{S}_2$ as input substrate, they must also provide a meaning for *presenting* $\mathbf{S}_1$ and $\mathbf{S}_2$ to the relevant constructor as the substrate $\mathbf{S}_1 \oplus \mathbf{S}_2$. Third, and most importantly, they must conform to Einstein's [28] *principle of locality* in the form:

II. There exists a mode of description such that the state of $\mathbf{S}_1 \oplus \mathbf{S}_2$ is the pair $(\xi, \zeta)$ of the states[5] $\xi$ of $\mathbf{S}_1$ and $\zeta$ of $\mathbf{S}_2$, and any construction undergone by $\mathbf{S}_1$ and not $\mathbf{S}_2$ can change only $\xi$ and not $\zeta$.

Unitary quantum theory satisfies II, as is explicit in the Heisenberg picture [28]. This, like many of the constructor-theoretic principles I shall be using, is tacitly assumed in all formulations of thermodynamics. Constructor theory states them explicitly, so that their physical content and consequences are exposed.

Tasks may be composed into networks to form other tasks, as follows. The *parallel composition* $\mathcal{A} \otimes \mathcal{B}$ of two tasks $\mathcal{A}$ and $\mathcal{B}$ is the task whose net effect on a composite system $\mathbf{M} \oplus \mathbf{N}$ is that of performing $\mathcal{A}$ on $\mathbf{M}$ and $\mathcal{B}$ on $\mathbf{N}$. When $\mathrm{Out}(\mathcal{A}) = \mathrm{In}(\mathcal{B})$, the *serial composition* $\mathcal{B}\mathcal{A}$ is the task whose net effect is that of performing $\mathcal{A}$ and then $\mathcal{B}$ on the same substrate. Parallel and serial composition must satisfy the *composition law*

III. The serial or parallel composition of possible tasks is a possible task,

---

[4] For how the appearance of stochasticity proper of quantum systems can be recovered in constructor theory, from purely deterministic, constructor-theoretic statements about possible/impossible tasks, see [35].

[5] In which case the same must hold for intrinsic attributes.

which is a tacit assumption both in information theory and thermodynamics, and finds an elegant expression in constructor theory. Note however that in constructor theory the composition of two impossible tasks may result in a possible one. In fact, this is one of the characteristics of the existence of a conservation law (section 4).

## 2.1. Possible tasks *vs* permitted processes

The second law of thermodynamics in constructor theory takes the form of what can be called a *law of impotence* [10]: a law requiring some task to be possible, and its transpose to be impossible. Such irreversibility, unlike that required by the second law in statistical mechanics, is compatible with time-reversal symmetric dynamical laws, as noted in [13]. The reason is rooted in the fundamental difference between a task being possible and a process being permitted, as follows.

I shall consider a physical system whose dynamics are expressible in the traditional conception, but which also conforms to constructor theory. Let the system's state space, containing all its states $\sigma$ described by the subsidiary theory be $\Gamma$. In the traditional conception [13], a process is represented as a trajectory – the sequence of states the system goes through as the evolution unfolds: $P \doteq \{\sigma_t \in \Gamma : t_i < t < t_f\}$. A process is *permitted* under the theory if it is a solution of the theory's equations of motion; let W be the set of all permitted processes. Let the map R transform each state $\sigma$ into its 'time-reverse' $R(\sigma)$. For example, R may reverse the sign of all momenta and magnetic fields. Also, define the time-reverse $P^-$ of a process P by: $P^- \doteq \{(R\sigma)_{-t} \in \Gamma : -t_f < t < -t_i\}$, [13]. The subsidiary theory is time-reversal invariant if the set W of permitted processes is closed under time reversal, i.e. if and only if: $P^- \in W \Leftrightarrow P \in W$.

Now consider a time-reversal invariant subsidiary theory and the constructor-theoretic statement that the task $\mathcal{T}$ is possible, but $\mathcal{T}^{\sim}$ is impossible. As we said, the second law (like any law of impotence) is expressed via a statement of this kind. We can immediately see that those two facts are compatible with one another:

That $\mathfrak{T}$ is possible implies that the process $P_\varepsilon$ corresponding to an approximation to a constructor $\vec{C}$ performing $\mathfrak{T}$ to accuracy $\varepsilon$, is permitted for any $\varepsilon$ (short of perfection). That the theory is time-reversal invariant implies that the process $P^-_\varepsilon$ too is permitted. But, crucially, $P^-_\varepsilon$ does not correspond to the task $\mathfrak{T}^\sim$ being performed to accuracy $\varepsilon$; because the reversed time-evolution of the approximate constructor running in the process $P_\varepsilon$ is not the dynamical evolution of an approximate constructor for $\mathfrak{T}^\sim$, to accuracy $\varepsilon$. Thus, the statement that $\mathfrak{T}$ is possible and $\mathfrak{T}^\sim$ is not possible is compatible with time-reversal invariant dynamical laws.

As explained in section 1, in constructor theory there is no need to require 'equilibrium states' to exist – defined as states that physical systems evolve spontaneously to, which never change thereafter unless the external conditions change. Existing formulations depend on this impossible requirement, via the so-called 'minus-first' law [16]. Here it will only be necessary to require there to be a particular class of intrinsic attributes, which I shall call *thermodynamic attributes*. Crucially, they need not have a definite temperature, and include many more than equilibrium states. Indeed thermodynamics in constructor theory is more general than standard equilibrium thermodynamics: the notion of temperature need never be invoked. Specifically, I shall require the following principle to hold:

IV. Attributes that are unchanged except when acted upon are possible.

Such attributes will be called 'thermodynamic attributes'. They generalise thermodynamic states, but they are fundamentally different from equilibrium states. In short, they are attributes of a physical system that can be stabilised to arbitrarily high accuracy (without side-effects); in the case of a qubit, they include quantum states that are very far from equilibrium: for instance, its pure states. The principle requires the possibility of bringing about such attributes to any accuracy short of perfection.

For example, consider a glass of water with temperature $x$. Principle IV requires that the attribute $x$ of the glass and water can be stabilised to any arbitrary accuracy, short of perfection. This is compatible with fluctuations occurring. For higher accuracies, stabilisation will require inserting various insulating materials (part of the constructor) around the glass of water to keep it at temperature $x$ to perform the task to that accuracy, while for lower accuracies just the glass by itself will suffice. So, principle IV is fundamentally different from the requirement that "equilibrium states exist". For while the former is not contradicted by fluctuations occurring, the latter is. Classical thermodynamics, relying on the latter requirement, is therefore scale-dependent, given the occurrence of fluctuations over sufficiently large time-scales. Constructor theoretic thermodynamics relies on the former, so its laws can be scale-independent.

One might wonder about the fact that constructors are just like equilibrium states, in that both never occur in reality. As I mentioned, however, constructor-theoretic laws are formulated exclusively in terms of possible/impossible tasks, *not* in terms of constructors. In particular, they never require perfect constructors to exist. Whenever requiring a task to be possible, one refers implicitly to a sequence of ever improving (but never perfect) approximations to the ideal constructor, with no limit on how each approximation can be improved. That there is, or there is not, a limit to how well the task can be performed, is a scale-independent statement.

As a consequence of laws being stated in terms of possible/impossible tasks, the emergence of an arrow of time and the second law of thermodynamics appear as entirely distinct issues [11, 30]. The former is about the spontaneous evolution of isolated physical systems, as established by, e.g., the minus-first law; the latter is about the possibility of certain tasks on finite subsystems of an isolated system. In line with axiomatic thermodynamics, I shall assume here the existence of an unambiguous 'before and after' in a physical transformation. This must be explained under constructor theory by subsidiary theories about time, in terms of constructor-theoretic interoperability laws (see section 3) concerning tasks of synchronising 'clocks' [31] – *not* in terms of the ordering established by minus-first law, as in current thermodynamics.

For present purposes, I shall restrict attention to subsidiary theories allowing for an unlimited number of substrates to be prepared in their thermodynamic attributes; and I shall require substrates to be *finite*. Having defined *generic substrates* as those substrates that occur in unlimited numbers [2], I shall assume that:

V.  The task of preparing any number of instances of any substrates with any one of its thermodynamic attributes from generic substrates is possible.

This is a working assumption about cosmology. It would be enough that it hold for a subclass of physical systems only, to which we confine attention for present purposes.

The constructor-theoretic concept of *side-effect*, which will be essential in understanding the notion of 'adiabatically possible' (section 5), is then introduced as follows: If $(\mathcal{A} \otimes \mathcal{C})^{\checkmark}$ for some task $\mathcal{C}$ on generic substrates (see [2]), $\mathcal{A}$ is possible with side-effects, which is written $\mathcal{A}^{\checkmark}$, and $\mathcal{C}$ is the side-effect, which occurs in the system's surroundings.

### 3. Constructor theory of information

I shall now summarise the principles of the *constructor theory of information* [2], which I shall use in sections 5 and 6 to define 'work media' and 'heat media', and to distinguish work from heat. They express the properties required of physical laws by the theories of (classical) information, computation and communication. Nothing that follows is probabilistic or 'subjective'; information is explained in terms of objective, *counterfactual* properties of substrates ('information media') – i.e. about what tasks are possible on them.
The logic is that one *first* defines *a class of substrates* as those on which certain tasks are possible/impossible. In the constructor theory of information these capture the properties of a physical system that would make it capable of instantiating what has been informally referred to as 'information'. Then, one expresses principles about them.

A *computation*[6] *medium* with *computation variable* $V$ (at least two of whose attributes have labels in a set $V$) is defined as a substrate on which the task $\Pi(V)$ of performing *every* permutation $\Pi$ defined via the labels $V$

$$\Pi(V) \doteq \bigcup_{x \in V}\{x \to \Pi(x)\}$$

is possible (with or without side-effects). $\Pi(V)$ defines a logically *reversible computation*.

*Information media* are computation media on which additional tasks are possible. Specifically, a variable $X$ is *clonable* if for some attribute $x_0$ of **S** the computation on the composite system $\mathbf{S} \oplus \mathbf{S}$

$$\bigcup_{x \in X}\{(x, x_0) \to (x, x)\}, \qquad (2)$$

namely *cloning X*, is possible (with or without side-effects)[7]. An *information medium* is a substrate with at least one clonable computation variable, called *information variable* (whose attributes are called *information attributes*). For instance, a qubit is a computation medium with *any* set of two pure states, even if they are not orthogonal [2]; it is an information medium with a set of two *orthogonal* states. Information media must also obey the principles of constructor information theory, which I now review:

**Interoperability of information.** Let $X_1$ and $X_2$ be variables of substrates $\mathbf{S}_1$ and $\mathbf{S}_2$ respectively, and $X_1 \times X_2$ be the variable of the composite substrate $\mathbf{S}_1 \oplus \mathbf{S}_2$ whose attributes are labelled by the ordered pair $(x, x') \in X_1 \times X_2$, where $X_1$ and $X_2$ are the sets of labels of $X_1$ and $X_2$ respectively, and $\times$ denotes the Cartesian product of sets. The *interoperability principle* is a constraint on the composite system of information media (and on their information variables):

---

[6] This is just a label for the physical systems with the given definition. Crucially, it entails no reliance on any a priori notion of computation (such as Turing-computability).

[7] The usual notion of cloning, as in the no-cloning theorem [32], is (2) with $X$ as the set of all attributes of **S**.

VI. The combination of two information media with information variables $X_1$ and $X_2$ is an information medium with information variable $X_1 \times X_2$.

This expresses the property that information can be copied from any one information medium to any other; which makes it possible to regard information media as a class of substrates. Interoperability laws for heat and work will be introduced in sections 5 and 6.

The concept of 'distinguishable' – which is used in the zeroth law (in section 6) – can be defined in constructor theory without circularities or ambiguities. A variable $X$ of a substrate **S** is *distinguishable* if

$$\left(\bigcup_{x \in X}\{x \to i_x\}\right)^{\checkmark} \qquad (3)$$

where $\{i_x\}$ is an information variable (whereby $i_x \cap i_{x'} = \emptyset$ if $x \neq x'$). I write $x \perp y$ if $\{x, y\}$ is a distinguishable variable. Information variables are necessarily distinguishable, by the interoperability principle VI. Note that 'distinguishable' in this context is not the negative of 'indistinguishable', as used in statistical mechanics to refer to bosons and fermions. Rather, it means that it is possible to construct a 'single-shot' machine that is capable of discriminating between any two attributes in the variable. For instance, any two non-orthogonal states of a quantum system are not distinguishable, in this sense.

I shall use the principle [2] that:

VII. A variable whose attributes are all pairwise distinguishable is distinguishable.

This is trivially true in quantum theory, for distinguishable pairs of attributes are orthogonal pairs of quantum states – however, it must be imposed for general subsidiary theories.

## 4. Conservation of energy

In constructor theory, conservation laws cannot be formulated via the usual dynamical considerations. As we shall see, the notion of a conserved quantity (and in particular energy) will refer to a particular pattern of possible/impossible tasks. To describe it, I shall now introduce a powerful constructor-theoretic tool – an *equivalence relation* ' $\simeq$ ' (pronounced *'is-like'*) on the set of all tasks on a substrate, which, ultimately, I shall use to express the intuitive notion that two tasks in the same equivalence class 'would violate the principle of the conservation of energy by the same amount'.

**'Is-like' equivalence relation.** Given any two *pairwise tasks* $\mathfrak{A} = \{x \to y\}$, $\mathfrak{B} = \{x' \to y'\}$, we say that $\mathfrak{A} \simeq \mathfrak{B}$ if and only if

$$[(\mathfrak{A}^\sim \otimes \mathfrak{B})^\checkmark \wedge (\mathfrak{A} \otimes \mathfrak{B}^\sim)^\checkmark].$$

This is an equivalence relation over the set of all pairwise tasks on *thermodynamic* attributes of substrates under a given subsidiary theory. So, the family of all equivalence classes generated by 'is-like' is a *partition* of that set. I shall assume initially that each class is labeled by a vector $\underline{\Delta} = (\Delta_i)$ of functions $\Delta_i : S_\mathbf{M} \Rightarrow \Re$, where $\Re$ is, for simplicity, the set of the real numbers, with the property that $\Delta_i(\mathfrak{A}) = \Delta_i(\mathfrak{B}), \forall i$ if and only if $\mathfrak{A} \simeq \mathfrak{B}$.

Another is-like relation on the set of thermodynamic attributes can also be defined, based on the serial composition of two tasks $\mathfrak{A}^\sim \mathfrak{B}$ and $\mathfrak{A}\mathfrak{B}^\sim$ (whenever it is defined, i.e., whenever $Out(\mathfrak{A}) = Out(\mathfrak{B})$):

$$\mathfrak{A} \stackrel{\bullet}{\simeq} \mathfrak{B} \leftrightarrow [(\mathfrak{A}^\sim \mathfrak{B})^\checkmark \wedge (\mathfrak{B}^\sim \mathfrak{A})^\checkmark]$$

One can easily prove that this, too, is an equivalence relation on the set of all pairwise tasks on thermodynamic attributes, that can be serially composed. By the composition law, $\mathfrak{A} \simeq \mathfrak{B} \Rightarrow \mathfrak{A} \stackrel{\bullet}{\simeq} \mathfrak{B}$ whenever $Out(\mathfrak{A}) = Out(\mathfrak{B})$ because: $\mathfrak{A} \simeq \mathfrak{B} \Rightarrow [(\mathfrak{A}^\sim \otimes \mathfrak{B})(\mathfrak{B} \otimes \mathfrak{B}^\sim)]^\checkmark \Rightarrow (\mathfrak{A}^\sim \mathfrak{B})^\checkmark$, and likewise for the transposes; conversely, $\mathfrak{A} \stackrel{\bullet}{\simeq} \mathfrak{B} \Rightarrow (\mathfrak{A}^\sim \mathfrak{B})^\checkmark \Rightarrow [(\mathfrak{A}^\sim \otimes \mathfrak{B})(\mathfrak{B} \otimes \mathfrak{B}^\sim)]^\checkmark \Rightarrow (\mathfrak{A}^\sim \otimes \mathfrak{B})^\checkmark$, where I have used the property that $[\mathfrak{T}\mathfrak{S}]^\checkmark \Rightarrow \mathfrak{T}^\checkmark$, where $\mathfrak{S}$ is the task of physically swapping two substrates – in the case above it is realised by the task $(\mathfrak{B} \otimes \mathfrak{B}^\sim)$. This is because the physical swap of two substrates is assumed to be a trivial tasks, just like the unit.

Hence, the partitions into equivalence classes generated by $\mathfrak{A} \simeq \mathfrak{B}$ and by $\mathfrak{A} \stackrel{\bullet}{\simeq} \mathfrak{B}$ are the same whenever they are both defined. Therefore they can be represented by the same labelling $\underline{\Delta}$, $\mathfrak{A} \simeq \mathfrak{B}$ being an extension of $\mathfrak{A} \stackrel{\cdot}{\simeq} \mathfrak{B}$ to the set of pairwise tasks on thermodynamic attributes.

The physical meaning of the partition into equivalence classes is that *tasks in the same class, if impossible, are impossible for the same reason*: for when $\mathfrak{A} \simeq \mathfrak{B}$ then $(\mathfrak{A}^\sim \otimes \mathfrak{B})^\checkmark$ and $(\mathfrak{A} \otimes \mathfrak{B}^\sim)^\checkmark$. For example, suppose in the traditional conception there is only one conservation law of some scalar (e.g. energy); in constructor theory, this corresponds to the tasks $\mathfrak{A}$ and $\mathfrak{B}$ that would both violate the conservation law by the same amount being in the same class. Both are impossible, but $(\mathfrak{A}^\sim \otimes \mathfrak{B})^\checkmark$ and $(\mathfrak{A} \otimes \mathfrak{B}^\sim)^\checkmark$ because the two tasks balance one another when performed in parallel, so that overall the conservation law is not violated. Similarly, consider a law of impotence, such as the second law, requiring tasks decreasing (or increasing) a given function (such as entropy) to be impossible, but their transpose to be possible. Then, is-like is such that tasks requiring a change in that function by the same amount belong to the same class.

Accordingly, one regards the equivalence classes' labels as the amount by which the conserved quantity (or the monotone such as entropy) is changed. To ensure that this identification is physically meaningful, one can show that under the additional constraints of locality (principle II), the functions $\Delta_i$ are additive, i.e., that for all *i*:

$$\Delta_i(\mathfrak{T}_1 \otimes \mathfrak{T}_2) = \Delta_i(\mathfrak{T}_1) + \Delta_i(\mathfrak{T}_2)$$
$$\Delta_i(\mathfrak{T}_1 \mathfrak{T}_2) = \Delta_i(\mathfrak{T}_1) + \Delta_i(\mathfrak{T}_2)$$

Thus, one assumes that for all *i* there exists an additive real-valued function $F_i : \Delta_i(\mathfrak{T}) = F_i(y) - F_i(x)$ – i.e., that $F_i((x,y)) = F_i(x) + F_i(y)$.

Now, it is easy to prove that there can only be *three kinds* of equivalence classes. One kind consists of only one class, containing the unit task $\mathfrak{I}$ and other possible tasks only, and it is labelled by $\Delta_i = 0, \forall i$. Given a representative element $\mathfrak{A}$ in that class, its transpose $\mathfrak{A}^\sim$ is in that class too, and it is possible.

Next is the kind of class with the property that each task $\mathcal{A}$ in the class is impossible, and its transpose, in the class labelled by $-\Delta_i(\mathcal{A}), \forall i$, is also impossible. A familiar example where these classes are non-empty is when there are tasks that violate a given conservation law by the same amount.

Finally, there is the kind of class with the property that each task in it is possible, but its transpose is impossible. Such a class reflects the presence of a "law of impotence", such as the second law of thermodynamics: tasks in the same class change entropy by the same amount. See the figure 2.

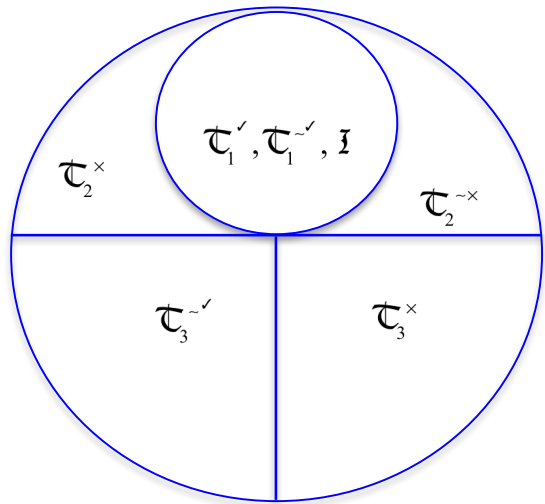
Figure 2

However, it is up to the subsidiary theory to say which one of those sectors in figure 2 is populated. For example, with no law of impotence, the first and second sector only would be populated. In thermodynamics it is customary to treat the simplified case that there is only one conservation law (namely that of energy), and one law of impotence (the second law). I shall do the same. Then there will be only two additive functions $F_1 = U(x), F_2 = \Sigma(x)$ such that $\underline{\Delta}(\mathcal{T}) = (\Delta U(\mathcal{T}), \Delta\Sigma(\mathcal{T}))$. In the first kind of class, containing the identity, $\Delta U = 0 = \Delta\Sigma$. In my notation, $U$ will be the function involved in the conservation law. At this stage $U$ could be any conserved quantity, not necessarily energy. It is energy only if it satisfies the first and second law of thermodynamics – see sections 5 and 6. (The second law, requiring that the classes with non-zero $\Delta\Sigma$ be non-empty, corresponds to requiring that there be what in classical thermodynamics are called 'thermodynamic coordinates', in addition to 'mechanical coordinates', i.e., coordinates that pertain to mechanical means only). In view of the above-mentioned simplification, I shall nevertheless call $U$ 'energy'.

The principle of **conservation of energy** can be then stated as follows:

VIII. The task of changing the $U$ of any substrate is impossible.

This conservation law requires the second kind of class (figure 2) to exist, containing tasks $\mathcal{T} = \{x \to y\}$ where $\Delta U(\mathcal{T}) \neq 0$, because, by VIII, $U(x) \neq U(y) \Rightarrow \{x \to y\}^{\times}, \{y \to x\}^{\times}$. For such tasks $\mathcal{T}$, $\Delta \Sigma(\mathcal{T})$ may or may not be zero. In the general case, the function $\Delta \Sigma$ will describe the class generated by a law of impotence – i.e., containing tasks $\mathcal{T}$ that are possible, but have an impossible transpose (the third kind in figure 2). For tasks in those classes it must be $\Delta U = 0$, because $\mathcal{T}$ is possible, and $\Delta \Sigma \neq 0$. Nothing in the principles so far requires the subsidiary theories to permit such tasks, but as we shall see, the second law of thermodynamics will precisely require them to do so; $\Delta \Sigma$ will then be connected to what we call entropy (see section 6).

Another interesting consequence of the difference between a task being *impossible* and a process being not permitted, is that a conservation law stated in these terms implies *that U of a substrate must be bounded both above and below* – a fact that must be otherwise imposed separately in the traditional conception (Deutsch, 2013). For, suppose that $U(y) - U(x) = \delta$ were not bounded above, and that there are no other reason why $\{x \to y\}$ is impossible. Then the task $(\mathcal{A}_\delta \otimes \{x \to y\})^{\checkmark}$ would be possible, for any task $\mathcal{A}_\delta$ : $\Delta U(\mathcal{A}_\delta) = \delta$, $\Delta \Sigma(\mathcal{A}_\delta) = 0$ because $\Delta(\mathcal{A}_\delta \otimes \{x \to y\}) = 0$. The first substrate, were $U$ not bounded, would still have the ability to perform the task $\{x \to y\}$, any number of times: it would therefore qualify as a constructor for the task $\{x \to y\}$, which would therefore be possible – contradicting the principle of conservation of energy. A similar contradiction follows from assuming that it is not bounded below. Thus, $U$ must be bounded above and below for *any substrate*.

Energy being bounded above is an unfamiliar requirement. But it is a consequence of the fact that the conservation law is about the energy of a *substrate* – i.e., a physical system that can be presented to a constructor. For there exists an energy value beyond which any physical system changes so drastically, for instance, by turning into a black hole or a vast plasma cloud – that no single (even idealised) constructor could accept it as input for arbitrarily large energies.

## 5. The first law of thermodynamics

I shall now recast the notion of *adiabatic accessibility* [4] in constructor theory. To this end I introduce a class of physical substrates, which I shall call *work media*. They generalise the 'weight in a gravitational field' that appeared in the original definition of adiabatic accessibility, and give a precise physical meaning to the notion of 'mechanical means', appearing in the definition of adiabatic enclosure (see section 1.2). The first law will then be stated as the requirement that such media be *interoperable* – i.e., that all ways of 'doing work' are interchangeable with one another.

The notion of work media is the core of this paper: it is key to the scale-independent formulation of adiabatic accessibility and of the second law. The meaning of the definition can be explained by noting that it addresses the following problem. Given a list of all possible/impossible pairwise tasks on the substrates of a subsidiary theory, the problem is to single out the substrates that behave like weights (i.e., the 'mechanical means', which can be used as side-effects of an adiabatic transformation) from those that behave like thermal states (thus displaying a 'heat-like' behaviour). The definition of work media gives a way to do that, at any scale. It is guided by the following physical intuition, holding in classical thermodynamics. There is *one property* that distinguishes substrates such as a weight at various heights from objects such as a glass of water at a range of temperatures. Such property is that a given task is *possible* on the former class of substrates, but it is impossible on the latter. Let the attributes in the variable $W = \{w_+, w_0, w_-\}$ denote three (ordered) different heights of the weight; and the attributes in $H = \{T_+, T_0, T_-\}$ denote three (ordered) different temperatures that the glass of water may assume. The task in question is the *swap* (defined below) of any two adjacent attributes of the substrate, *with the only allowed side-effect* being that of affecting a replica *of the same substrate*, initialised *in one of the two attributes in question*. Indeed, for any two attributes $w_+, w_0 \in W$ their swap is *possible* on two replicas of the weight:

$$\{(w_+, w_0) \to (w_0, w_+), (w_0, w_0) \to (w_+, w_-)\}^{\checkmark}$$

but it is impossible on two replicas of a glass of water:

$$\{(T_+, T_0) \to (T_0, T_+), (T_0, T_0) \to (T_+, T_-)\}^{x}$$

This is because the latter task would violate the second law of thermodynamics (by requiring that a substrate with uniform temperature be changed to one with a temperature difference, *with no other side-effect allowed*). This crucial difference between those two kinds of substrates is a *counterfactual* property.

On this ground, I shall now give the general definition of work media.

**Work Media.** A *work medium* **M** with *work variable* $W = \{w_1, w_2, ..., w_N\}$ is a substrate whose thermodynamic attributes $\{w_1, w_2, ..., w_N\}$ have the property that:

1. $\{w_i \to w_j\}^{x}$ for all $i, j$.

2. For any *pair of adjacent attributes* $\{w_n, w_{n+1}\} \subseteq W$:

    a. $\{w_n \to w_{n+1}\} \simeq \{w'_n \to w'_{n+1}\}$ for all adjacent pairs of attributes $\{w'_n, w'_{n+1}\} \subseteq W$;

    b. For *some* attributes $x, x' \in W$ and, crucially, $x_0 \in \{w_n, w_{n+1}\}$:

    $$\{(w_n, x_0) \to (w_{n+1}, x), (w_{n+1}, x_0) \to (w_n, x')\}^{\checkmark}$$

c. If $w_n = (a_n, b_n), w_{n+1} = (a_{n+1}, b_{n+1})$ for some thermodynamic attributes $\{a_n, b_n, a_{n+1}, b_{n+1}\}$, then the variables $\{a_n, a_{n+1}\}$, $\{b_n, b_{n+1}\}$ separately satisfy all the above conditions.

*Condition 1* requires that $\Delta U(\{\mathcal{T}_{i,j}\}) \neq 0, \forall i,j$: for, by the conservation law VIII (under the simplifying assumption) a necessary condition for both a task and its transpose to be impossible is that they would change the $U$ of their substrate.

*Condition 2a* implies that adjacent attributes in a work variable are "equally spaced": $\Delta U(\{\mathcal{T}_{i,j}\}) = \Delta$ for all adjacent $i,j$. Whenever $\Delta U(\mathcal{T}) = \Delta U(\mathcal{T}')$ for some *pairwise* tasks $\mathcal{T}$ and $\mathcal{T}'$ over some two-fold variables $V$ and $V'$, I shall say that $V$ is *U-commensurable* with $V'$, and that $\mathcal{T}$ is *U-commensurable with* $\mathcal{T}'$.

*Condition 2b* is the key counterfactual property that singles out substrates such as weights. It requires that it is possible to perform a *swap* of any two adjacent attributes $\{w_n, w_{n+1}\}$ of **M** with a side-effect on a replica of the same substrate, initialised so that $W$ is sharp with the same value – i.e., the substrate holds either a sharp $w_n$ or a sharp $w_{n+1}$. I shall call *'work attributes'* the attributes in the variable $W' \subset W$ with the property that for any two adjacent attribute $w_n, w_{n+1} \in W'$, property 2b is satisfied with *both* $x_0 = w_n$ and $x_0 = w_{n+1}$.

The scale-independent, more general notion of a work medium generalises that of 'mechanical means', and it is consistent with existing thermodynamics. This can be explained as follows.

Clearly, a quantum system with any number greater than 3 of equally spaced energy levels does satisfy the definition of work media: its work attributes are all except the ones labeled by extremal energy values. For instance, a weight in a gravitational field, with three distinct 'heights', is a work medium.[8] As explained above, classical thermodynamics offers also a key example of substrates that do *not* satisfy that condition for work media. Consider a quantum system with a variable containing different *thermal states*. Consider any two-fold subset of that variable: say $\{\rho_1, \rho_2\}$ – these correspond to two macrostates of the system, with different temperatures 1 and 2. The reason why this is not a work medium is that, as we said:

$$\left\{ (\rho_1, \rho) \to (\rho_2, \rho_x), (\rho_2, \rho) \to (\rho_1, \rho_y) \right\}^{\times}, \forall \rho \in \{\rho_1, \rho_2\}$$

(where $\rho_x, \rho_y$ are allowed to be any two other quantum (pure or mixed) states with different mean[9] energies from the initial ones, as required by conservation of energy). Whatever the state $\rho$ may be between $\rho_1, \rho_2$, the *swap* with that constrained side-effect, plus the conservation of energy, would require a thermal state with a given temperature, $(\rho, \rho)$, to be changed to one where there is a temperature difference, such as $(\rho, \rho_x)$ – which is forbidden by the second law in classical thermodynamics. This is a crucial difference between a 'mechanical system' (e.g. a weight in a gravitational field), and one having thermal states (e.g. a reservoir with a range of possible temperatures). This difference can be stated in a scale-independent way *only* by expressing the counterfactual properties of that system.

It is possible to show that the only stationary quantum states of a single quantum system (i.e., diagonal in the energy basis) that qualify as work attributes of a work variable are pure eigenstates of the unperturbed

---

[8] Work media necessarily have discrete work variables. Continuous spectra can be approximated to arbitrary accuracy by composing a number of such physical systems.

[9] A quantum thermal state with a given mean energy corresponds, in constructor theory, to a thermodynamic attribute (i.e., one that can be stabilised to arbitrary accuracy – *not* the result of some spontaneous thermalisation process of an isolated system.)

Hamiltonian of the system. Thus, the above definition of work media is consistent with existing thermodynamics: it is a good candidate to provide foundations for the new definition of adiabatic accessibility.

*Condition 2c* (mirroring the requirement that there are no other conservation laws) rules out from the class of work variables of composite systems those whose attributes are such that when transforming one into another the change in $U$ is not the only one taking place. For, in the presence of a law of impotence, a task and its transpose could both be impossible, but a change in the other label of the equivalence classes, $\Sigma$, might take place too. For instance, the variable of the composite system $\{(U_+,\Sigma_+),(U_0,\Sigma_0),(U_-,\Sigma_-)\}$, where U can be thought of as being internal energy and $\Sigma$ can be thought of as being classical thermodynamics' entropy, would satisfy conditions 1-2.b. However, it does not satisfy conditions 1-2c, for $\{U_+,U_0,U_-\}$ satisfies properties 1-2b while $\{\Sigma_+,\Sigma_0,\Sigma_-\}$ does not satisfy property 2a. It follows immediately that the minimal work variable in the presence of a single conservation law is one that has *three* thermodynamic equally spaced attributes, labelled by different values of $U$. [10]

Note that a perfect work medium need not exist in reality: it is enough that they be approximated arbitrarily well. Such a medium might be made, for instance, as a composite system of several systems with energy spectra that are *not* equally spaced.

**Work variables are information variables.** Condition 2b – requiring what I shall call the 'swap' property – provides an unexpected, illuminating connection between thermodynamics and information theory: any sub-variable $\{w_n, w_{n+1}\}$ of a work variable is *distinguishable*, in the exact, constructor-theoretic sense of section 3. To see how, recall that [2] any two disjoint intrinsic attributes $x$ and $x'$ are *ensemble distinguishable*, which means the following.

---

[10] The class of work media is more general than that of 'mechanical systems' given in [34]; the latter is based on the notion of reversibility of spontaneous processes on isolated systems; while the former on possible/impossible tasks.

Let $\mathbf{S}^{(n)}$ denote a physical system $\overbrace{\mathbf{S} \oplus \mathbf{S} \oplus \ldots \mathbf{S}}^{n \text{ instances}}$ consisting of $n$ instances of $\mathbf{S}$, and $x^{(n)}$ the attribute $\overbrace{(x,x,\ldots x)}^{n \text{ terms}}$ of $\mathbf{S}^{(n)}$. Denote by $\mathbf{S}^{(\infty)}$ an unlimited supply of instances of $\mathbf{S}$. This is of course a theoretical construct, which does not occur in reality. That $x$ and $x'$ are ensemble-distinguishable means that the attributes $x^{(\infty)}$ and $x'^{(\infty)}$ of $\mathbf{S}^{(\infty)}$ are distinguishable. In quantum theory, this corresponds to the fact that any two different quantum states are asymptotically distinguishable – a property at the heart of so-called quantum tomography. Now, let $z \in \{w_n, w_{n+1}\} \subseteq W$, where $W$ is a work variable of some work medium. By property 2b it is possible to apply the *swap* operation to the work medium any number of times and the output would be a composite work medium $\mathbf{M} \oplus \mathbf{M} \oplus \mathbf{M} \ldots$ with the attribute $(w_{n+1}, x, x', x \ldots)$ if $z = w_n$; and with the attribute $(w_n, x', x, x', \ldots)$ if $z = w_{n+1}$. Thus, preparing the attributes $x^{(\infty)}$ and $x'^{(\infty)}$ would be a possible task, because of the assumption V about unbounded number of thermodynamic attributes being preparable from generic resources. Those attributes can be constructed to arbitrarily high accuracy, short of perfection, using a finite number of substrates for each accuracy. The attributes $x$ and $x'$ are intrinsic thermodynamic attributes, so that $x^{(\infty)} \perp x'^{(\infty)}$. Thus, by preparing $z^{(\infty)}$ from $\{w_n, w_{n+1}\}$ one could distinguish $w_n$ from $w_{n+1}$. From this 'pairwise' distinguishability of its attributes it follows, via the principle VII, that a work variable is a distinguishable variable. Thus, it is an information variable [2]. Hence all *work media are information media* – with their work variable being an information variable.

In general, that the swap on a pairwise computation variable such as $\{w_n, w_{n+1}\}$ is possible need not imply that the variable is distinguishable. For instance, any two non-orthogonal quantum states can be swapped [2]. It is the presence of a conservation law that requires an ancilla to perform a computation on attributes labeled by different values of the conserved quantity. A record of the state being swapped is left in the ancilla, whereby it is possible to distinguish the attributes in the work variable. This is the origin of the connection between a conservation law and information theory.

**Interoperability of work media.** The first law of thermodynamics in the traditional approach can be thought of as requiring that all ways of doing work are equivalent, and interchangeable, with one another. Constructor theory generalises this idea: the first law requires that work media are *interoperable* with one another – i.e., that there be a unique class of work media – as follows:

**I.** Given any two work media $M_1, M_2$ with commensurable work variables $W^{(1)}$, $W^{(2)}$:

  i. For any adjacent pair $\{w, w'\} \subseteq W^{(1)}$

  $$\{(w, x_0) \to (w', x), (w', x_0) \to (w, x')\}^{\checkmark} \qquad (4)$$

  for some attributes $x, x' \in W^{(2)}$ and $x_0 \in \{w, w'\} \subseteq W^{(1)}$; and likewise when the labels 1 and 2 are interchanged.

  ii. The composite substrate $M_1 \oplus M_2$ is a work medium, with variable $W \subseteq W^{(1)} \times W^{(2)}$, where $W$ is obtained by a relabeling of $W^{(1)} \times W^{(2)}$ where attributes with the same $U$ are assigned the same label.

Property (i) implies immediately that pairwise tasks defined on any two commensurable work variables are like one another, and that the task of transforming any two attributes in $W \subseteq W^{(1)} \times W^{(2)}$ having the same energy $U$ into one another is possible. From now on, given a work medium $M_1$, any substrate $M_2$ that satisfies equation (4) in property (i) above I shall call a *work-like ancilla* for $M_1$. Property (ii) is the *interoperability law for work media* – it requires the composite system of any two work media to be a work medium, which captures the intuition that a single battery can be substituted for two batteries of appropriate capacity. Two different work media with commensurable variables, such as a weight and a spring, are *interchangeable* with one another in this sense.

**Quasi-work media.** Work media are special kinds of substrates, to be used to generalise the notion of adiabatic accessibility. However, they do not exhaust all physical systems that in classical thermodynamics we would characterise as mechanical means. For instance, in the latter category we might well include a quantum system with exactly two energy levels, but the latter would not qualify as a work medium (because it does not allow for the swap property 2.b). Thus, it is useful to introduce a class of closely related substrates, *quasi-work media*, any pair of whose attributes can be swapped via a side-effect on a work variable; but which need not be usable as work-like ancillas – i.e., they need not be usable as a side-effect for the swap (condition 2b) that defines work media. As we shall see, swapping their attributes in that way *only* changes their $U$ (and not the possible other labels related to a law of impotence) – so, they too can be characterised as having mechanical coordinates only.

A *quasi-work medium* is a substrate with thermodynamic attributes $\{x,y\}$ (called *quasi-heat attributes*) such that $\{x \to y\} \simeq \{w \to w'\}$ for some work variable $\{w,w'\}$. All work media are quasi-work media with any pair of attributes in their work variables. Conversely, two-level system qualifies as a quasi-work medium, but not, as we said, as a work-medium. (Table 1 summarises these notions).

**Adiabatic possibility.** I now generalise the notion of 'adiabatic accessibility' in constructor-theoretic terms. The task $\mathfrak{T} = \{x \to y\}$ is *adiabatically possible* (denoted by $\{x \to y\}^{\checkmark}$) if it is possible with a side-effect task *over work variables only*:

$$\{\{x \to y\} \otimes \{w_1 \to w_2\}\}^{\checkmark}.$$

Vice-versa, it is *adiabatically impossible*, which I denote by $\{x \to y\}^{\stackrel{x}{=}}$, if

$$\left\{\left\{x \to y\right\} \otimes \left\{w_1 \to w_2\right\}\right\}^{x}$$

for all work variables $\left\{w_1, w_2\right\}$. Hence, whenever a task is possible it is also adiabatically possible, with a trivial side-effect (consisting of the unit task on some work attribute).

This definition is consistent with Lieb & Yngvason's [4], and makes the latter more general and scale-independent, for work media are defined via statements about possible/impossible tasks only.

Because of interoperability of work media, it is immediate that $\mathcal{T}^{\checkmark}$ and $(\mathcal{T}^{\sim})^{\checkmark}$ for any task $\mathcal{T} = \left\{w_1 \to w_2\right\}$ whose input/output attributes belong to a work variable, or to a quasi-work variable (but not necessarily for a work-like ancilla). The second law will require the existence of tasks for which $\mathcal{T}^{\checkmark}$, but $\mathcal{T}^{x}$ – and hence, a new class of substrates.

## 6. The second law of thermodynamics

The new notion of *adiabatic possibility* can now be used to state the second law of thermodynamics in a scale-independent way. I shall first introduce the notion of 'heat', via the powerful constructor-theoretic method of defining a class of substrates with certain counterfactual properties, obeying another interoperability law. This definition is, unlike that of classical thermodynamics, based on scale-independent statements about possible/impossible tasks. To that end, one needs an auxiliary class of substrates: quasi-heat media.

**Quasi-heat media.** A *quasi-heat medium* is a substrate with a variable $Q$ whose thermodynamics attributes have the property that $\forall \{h, h'\} \subseteq Q$, $\{h \to h'\}^{\checkmark}$, but $\{h' \to h\}^{x}$. $Q$ is called a *quasi-heat variable*.

A quasi-heat medium is not a work medium. If it were, then it would be possible to swap its heat attributes adiabatically, by the first law of thermodynamics (property (i)), contrary to assumption. Likewise, it is impossible for a quasi-heat medium to be a quasi-work medium.

The second law requires the subsidiary theory to permit a particular kind of quasi-heat media, as we shall see. To introduce them it is helpful to define, given a variable $H$, the *symmetric variable* $H' \subseteq H \times H$ of the composite substrate $\mathbf{M} \oplus \mathbf{M}$ with the property that its attributes are invariant under the swapping the two substrates. So, for instance, if $H = \{h, h'\}$, then $H' = \{(h,h), (h',h')\}$.

**Heat Media.** A *heat medium* with *heat variable* $H$ is a quasi-heat medium $\mathbf{M}$ whose quasi-heat variable $H$ has the following additional properties:

1. Each attribute $h \in H$ is *not distinguishable* from any other thermodynamic attribute.
2. $\mathbf{M} \oplus \mathbf{M}$ is a quasi-heat medium with the symmetric variable $H' \subseteq H \times H$
3. For any pair of attributes $\{h, h'\} \subset H$ there exists $h_F \in H$ such that $\mathbf{M} \oplus \mathbf{M}$ is a *quasi-work medium* with variable $\{(h',h),(h_F,h_F)\}$ (i.e., $\{(h',h) \to (h_F,h_F)\} \simeq \mathfrak{W}$ for some work variable $\mathfrak{W}$).
4. There is no work medium $\mathbf{W}$ such that $\mathbf{M}$ is a work-like ancilla for $\mathbf{W}$.

For consistency with existing thermodynamics, note that thermal states would satisfy all of these conditions. They satisfy **Condition 1**: since they have support on the set of all eigenstates of energy of a quantum state, they are not perfectly distinguishable by a 'single-shot' measurement from any other quantum state, and thus are not distinguishable according to the constructor-theoretic definition [11] (section 3). **Condition 2**, combined with the

---

[11] For instance, a thermometer cannot discriminate to arbitrarily high accuracy two quantum states with different Boltzmann distributions (i.e. with two different temperatures), for they are not

interoperability law for heat media (see below), is a generalisation of the property that it is impossible to 'extract work' from thermal states, no matter how many replicas of a substrates are available (this is also called *complete passivity* [32]). **Condition 3**, conversely, generalises the requirement that there be ways of 'extracting work' from certain other states of a composite substrates involving thermal states. For instance, $(h,h')$ could be the attribute of a composite substrate of two finite heat reservoirs, at different temperatures $h,h'$, with the same heat capacity, in which case the attribute $(h_F,h_F)$ is the attribute of those two reservoirs having the same temperature $h_F = \sqrt{hh'}$. It is well known that it is possible to transform the substrates made of the two reservoirs from the former attribute to the latter and vice-versa, adiabatically, by running a "reversible" heat engine (one that has no net change in entropy), or a refrigerator, between the two reservoirs. The adiabatic side-effect would be, respectively, a weight being raised (to absorb the work done by the heat engine in depleting the temperature difference in $(h,h')$); and a weight being lowered to power the refrigerator which can transform $(h_F,h_F)$ to $(h,h')$. Constructor theory gives us an interesting new insight: because of the requirement that $h \not\perp h'$, the task

$$\{(h,h',w_0) \to (h_F,h_F,w), (h_F,h_F,w_0) \to (h,h',w')\}$$

must be impossible for any work attributes $w_0$, $w$, $w'$. If it were possible, then $h \perp h'$, (because work variables are distinguishable) – contrary to assumption. Thus, although there can be constructors that perform each task separately, as we said, it is impossible to have a constructor that would perform their union.

The meaning of **condition 4** becomes clear when one examines the substrates it rules out. Let us consider a heat medium **M**, with the attributes $\{h_+,h_0,h_-\} \subset H$, that violates it – i.e., it is a work-like ancilla. Then, for some

---

orthogonal states. In constructor theory, this fact is shown not to be accidental, but essential to the nature of heat.

work medium **M** with some work variable $W \supseteq \{w_+, w_0, w_-\}$ (such that attributes with the same suffix have the same energy $U$):

$$\{(w_+, h_0) \to (w_0, h_+), (w_0, h_0) \to (w_+, h_-)\}^{\checkmark}$$

which would imply that $\{(w_+, h_0) \to (w_0, h_+)\}^{\checkmark}, \{(w_0, h_0) \to (w_+, h_-)\}^{\checkmark}$.

Thus, condition 4 rules out, from the logically possible assignments of adiabatically possible/impossible tasks on any 3-fold heat variable that are compatible with the definition of a heat medium, that which would require those two tasks to be possible (see figure 3, case (b)).

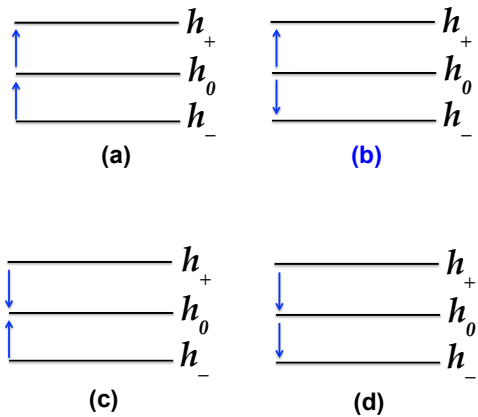

**Figure 3** Possible assignments of adiabatically possible tasks on a heat medium whose attributes (+,0,-) are labelled by decreasing values of energy. (The arrows represent adiabatically possible tasks, their transpose (not represented) is adiabatically impossible by definition of heat media.)

This property of heat media is central in deriving Kelvin's statement of the second law (section 7); and it is is compatible with properties of thermal states in traditional thermodynamics. Note that in classical thermodynamics both cases (b) and (c) in figure 3 are ruled out by an *ancillary law* [4, 10, 11] requiring that all adiabatically possible tasks whose transpose is adiabatically impossible either always increase (case (a)) or always decrease (case (d)) the energy. In constructor theory, that law will not be necessary.

**Interoperability of heat.** Heat media will be required to be a class of interchangeable substrates – again, by an interoperability law. The *interoperability law for heat media* requires that:

IX. The composite system of any two heat media with variables $H_1, H_2$ whose attributes have the same is-like labels, is a heat medium with heat variable $H' \subseteq H_1 \times H_2$.

This new principle requires the composite substrate of *any* two heat media, whatever their physical details, still satisfies properties 1–4 above. This is tacitly assumed in classical thermodynamics, but has this elegant expression in constructor-theoretic terms. In the traditional conception's approach to thermodynamics one usually demonstrates, given a particular subsidiary theory, that there exist models of physical systems displaying properties such as those required by the conditions 1-4. Here, instead, the logic is to express those properties in a scale-independent way, and then illustrate the principles that substrates which classify as heat, work, quasi-work and quasi-heat media must obey.

From now on, whenever a subsidiary theory requires a pattern of possible/impossible tasks on thermodynamic attributes conforming to there being a heat medium with heat variable *H*, I shall say that 'A heat medium/variable *H* is *mandated*'. Likewise for the other categories introduced so far: work, quasi-work and quasi-heat media. See table 1 for a summary of the intuitive meaning of each of those categories.

> *Work media* are objects like *weights*: they must have at least three distinguishable attributes with different energy, characterised by a particular 'swap' property.
> **Quasi-work media** include work media, but they need not have that swap property – e.g. a quantum system with only two energy states. The task of changing any of their attributes into any other is adiabatically possible.
> A **work-like ancilla** is a substrate with at least three thermodynamic attributes that can be used as a side-effect to swap work attributes of another system; but the task of changing any one of its attributes into any other may be adiabatically impossible.
> A **quasi-heat medium** is a substrate with at least a pair of attributes such that the task of changing one into another is adiabatically impossible, but has an adiabatically possible transpose.
> *Heat media* are quasi-heat media with at least three thermodynamic attributes that are not distinguishable from any other attribute and have certain additional properties (e.g. a quantum system with three different temperature states).

**Table 1**: *Informal description of substrates appearing in constructor-theoretic thermodynamics. (Terms in **boldface italic** denote media with direct analogues in classical thermodynamics.)*

**The second law of thermodynamics.** I shall now state the second law of thermodynamics, in scale-independent terms:

**II.** Consider any two attributes $x$ and $y$ in any two work variables of the same medium. That substrate is also a heat medium with a heat variable $H$ such that it contains a pair of heat attributes with the same energies as $x$ and $y$.

This is the constructor-theoretic generalisation of the second law. It requires that whenever a subsidiary theory mandates a work medium (whose attributes, recall, can be swapped only by changing their mechanical coordinates, i.e., with side-effects on work media only) – then given *any* pair of attributes in its work variables (including pairs with the same energy) it must also mandate a pair of heat attributes in a heat variable with the same energies. Thus, for any pairwise task on a pair of attributes belonging to work variables (which, by definition, must be such that either both it and its transpose are possible, or they are both impossible) there is a corresponding *U-commensurable* task on a pair of heat attributes (which, by definition of heat media, must be adiabatically possible, and its transpose adiabatically impossible; or vice-versa). This is reminiscent of Caratheodory's notion of there being adiabatically inaccessible points in any 'neighbourhood' of any point in the thermodynamic space. However, it does require the set of thermodynamic attributes to be a continuum.

Note that no notion of entropy nor temperature has been mentioned so far: this is *not* equilibrium thermodynamics. To the end of introducing an equivalent of *entropy*, let me now define another equivalence relation.

**Adiabatic is-like.** We say that $\mathfrak{A} \cong \mathfrak{B}$ (read: $\mathfrak{A}$ 'is adiabatically-like' $\mathfrak{B}$) if and only if

$$[(\mathfrak{A}^\sim \otimes \mathfrak{B})^{\checkmark} \wedge (\mathfrak{A} \otimes \mathfrak{B}^\sim)^{\checkmark}].$$

Once again, this is an equivalence relation on the set of all pairwise tasks on thermodynamic attributes. Under the simplifying assumption of a *single conservation* law and *a single law of impotence* (see section 5), the case where both $\mathcal{T}$ and its transpose are adiabatically impossible does not occur.[12] Consequently, instead of a vector of labels it is enough for present purposes that there is a single real-valued, additive function $\Delta_A(\mathcal{T})$ labeling the different classes, with the property that $\Delta_A(\mathcal{T}) = 0 \Leftrightarrow \mathcal{T}^{\checkmark} \wedge (\mathcal{T}^{\sim})^{\checkmark}$. Let $\mathcal{T} = \{x \to y\}$. By locality, I shall assume that $\Delta_A(\mathcal{T}) = \Delta S(\mathcal{T}) = S(y) - S(x)$ where $S$ is an additive, real-valued function whose domain is the set of thermodynamic attributes of all substrates.

**Entropy.** Now, one can show that the function $S$ has the following properties, which allow one to identify it as the constructor-theoretic generalisation of traditional entropy:

(i) $\Delta S(\mathcal{T}) = 0 \Leftrightarrow \mathcal{T}^{\checkmark} \wedge (\mathcal{T}^{\sim})^{\checkmark} \Leftrightarrow \mathcal{T} \cong \mathcal{I}$.

Therefore, any task $\mathcal{T} = \{w_1 \to w_2\}$ whose input/output attributes are work attributes belongs to the class labeled by $\Delta S(\mathcal{T}) = 0$.

In addition, one can show that the function $\Sigma$ introduced in the is-like relation (section 4) is related to S, as follows:

(ii) $\Delta \Sigma(\mathcal{T}) \neq 0 \Rightarrow \Delta S(\mathcal{T}) \neq 0$.

This is because:

$$\mathcal{T}^{\checkmark} \Rightarrow \mathcal{T}^{\checkmark} \wedge \Delta U(\mathcal{T}) = 0 = \Delta U(\mathcal{T}^{\sim})$$
$$\mathcal{T}^{\times} \Rightarrow \mathcal{T}^{\sim \checkmark} \vee \mathcal{T}^{\sim \times}.$$

The first line follows from the fact that whenever a task is possible, it is also adiabatically possible (the side-effect task being the unit task). The second line follows from the assumption that there is only one law of impotence. However, the option $\mathcal{T}^{\sim \checkmark}$ is not viable. For that would imply that either $(\mathfrak{T}^\sim)^\checkmark$ (which contradicts the premises) or that $(\mathfrak{T}^\sim \otimes \{w_1 \to w_2\})^\checkmark$ for some work variable $\{w_1, w_2\}$. However, this would require that $\Delta U(\mathfrak{T}^\sim) \neq 0$, by additivity of $U$, contrary to the assumptions. This proves (ii).

---

[12] This simplifying assumption implies that if a task $\mathcal{T}$ is adiabatically impossible, then its transpose is adiabatically possible. This property is essentially the *comparability axiom* of Lieb & Yngvason [4], but I shall not require it in this treatment.

As a consequence of (i), $\forall \mathfrak{T} : \mathfrak{T}^{\checkmark} \wedge (\mathfrak{T}^\sim)^{x}$, $\Delta S(\mathfrak{T}) \neq 0$. For, by property (ii):

$$(\mathfrak{T} \otimes \mathfrak{w})^\checkmark \Rightarrow \Delta U(\mathfrak{T} \otimes \mathfrak{w}) = 0 = \Delta U((\mathfrak{T} \otimes \mathfrak{w})^\sim)$$
$$(\mathfrak{T}^\sim \otimes \mathfrak{w})^x \wedge \Delta U((\mathfrak{T} \otimes \mathfrak{w})^\sim) = 0, \forall \mathfrak{w} \Rightarrow \Delta \Sigma(\mathfrak{T} \otimes \mathfrak{w}) \neq 0 \Rightarrow \Delta S(\mathfrak{T} \otimes \mathfrak{w}) \neq 0$$

Since $\Delta S(\mathfrak{w}) = 0$, by additivity: $\Delta S(\mathfrak{T}) \neq 0$.

Thus, whenever a task $\mathfrak{T}$ is adiabatically possible, but its transpose is not, $\Delta S(\mathfrak{T}) \neq 0$; we can therefore identify it as the constructor-theoretic generalisation of entropy. Assuming additional requirements, e.g. about the continuity of the space of thermodynamic points, allows one to show that the function has the property that whenever a task $\mathfrak{T}$ is adiabatically possible, but its transpose is not, then $\Delta S(\mathfrak{T}) > 0$. However, unlike in those cases, in constructor theory the physical content of the second law resides in the notion of adiabatic possibility, *not* in the properties of the entropy function. Thus, for present purposes, I shall not assume any of those additional requirements. Because of property (ii), the partition into equivalence classes generated by $S$ refines that generated by $\Sigma$. Therefore, one can uniquely identify a class generated by both the is-like and the adiabatic is-like equivalence relation by the labels $\underline{\Delta}(\mathfrak{T}) = (\Delta U(\mathfrak{T}), \Delta S(\mathfrak{T}))$ where $U$ is the function defined via the is-like relation and $S$ is that defined by the adiabatic-is-like relation. The combination of the first and second law imply that, if there are possible and impossible tasks at all, then *all* kinds (see figure 2) of the is-like equivalence relation are present.

**The Zeroth Law of Thermodynamics.** The *zeroth law* was an 'afterthought' in classical thermodynamics [26]: it was proposed, historically, after all the others, to introduce the notion of temperature. In constructor theory it is, too. However, its implications are somewhat different – in particular, it does not require the existence of temperature; nor is it about equilibration. The zeroth law is stated as follows:

X. Given any thermodynamic attribute $x$, $\{x \rightarrow h\}^{\checkmark}$ for any heat attribute $h$ having the same energy as $x$.

This requires that it is possible to convert any thermodynamic attribute into a heat attribute – i.e., one that cannot be distinguished from any other attribute. An example of such task might be that of converting some amount of purely mechanical energy completely into heat by 'rubbing'. However, note that this law differs both in form and content from existing ancillary laws in classical thermodynamics [10]: it is *not* about spontaneous processes occurring (i.e., equilibration), but about the possibility of a task; in addition, it does not require there to be a definite sign for the change in energy accompanying an adiabatically possible task with an adiabatically impossible transpose.

**7. Kelvin's statement of the second law**

For consistency with the traditional formulation of thermodynamics, I shall now prove that the constructor-theoretic version of Kelvin's statement, that *"heat cannot be converted entirely into work by means of a thermodynamic cycle"*, follows from the laws of thermodynamics expressed above, in the form:

The task $\{h \rightarrow w\}$ is impossible for every work attribute $w$ and heat attribute $h$.

For, suppose that task were possible for some such attributes: $\{h_0 \rightarrow w_0\}^{\checkmark}$ with $U(h_0) = U(w_0)$. Consider a 3-fold work variable $\{w_+, w_0, w_-\}$ including $w_0$ (which must exist by definition of work attribute, section 5). By the second law there must be a heat variable H with sub-variable $\{h_+, h_0, h_-\} \subseteq H$ with $U(h_+) = U(w_+)$ and $U(h_-) = U(w_-)$.

Now, recall, there are three possible ways of assigning adiabatic possibility/impossibility to any three-fold sub-variable $\{h_+, h_0, h_-\}$ of a heat variable (figure 3). In all such cases a contradiction is reached. Suppose that $\{h_0 \to h_-\}^{\underline{x}} \wedge \{h_+ \to h_0\}^{\underline{x}}$ (case (a) in figure 3). Then, by the interoperability of work (second line) and by the zeroth law (third line):

$$\{(h_0, w_0) \to (w_0, w_0)\}^{\checkmark}$$
$$\{(w_0, w_0) \to (w_-, w_+)\}^{\checkmark}$$
$$\{(w_-, w_+) \to (h_-, w_+)\}^{\checkmark}$$

By definition of adiabatically possible:

$$\{(h_0, w_0) \to (h_-, w_+)\}^{\checkmark} \Rightarrow \{h_0 \to h_-\}^{\underline{\checkmark}}$$

which violates the definition of heat medium, for this means that two of its heat attributes are adiabatically accessible from one another. The same line of argument leads to a contradiction when $\{h_0 \to h_+\}^{\underline{x}} \wedge \{h_- \to h_0\}^{\underline{x}}$ (case (d)) and when $\{h_0 \to h_+\}^{\underline{x}} \wedge \{h_0 \to h_-\}^{\underline{x}}$ (case (c)). Therefore the laws of constructor-theoretic thermodynamics require that heat cannot be completely converted into work.

This constitutes the promised unification of the axiomatic approaches with and Kelvin's. The principles used to achieve this result do not require a definite sign of the change in $U$ for adiabatically possible tasks with an adiabatically impossible transpose. Thus, the recovery of Kelvin's statement rests on different physical laws from the ancillary laws invoked by, e.g., [4, 10].

Only at this stage can one introduce, without circularities, the notions of 'doing work on', and 'transferring heat to', a substrate.

In any given construction to perform the task $\mathfrak{T} = \{x \to y\}$ on a substrate **M** such that $\Delta S(\mathfrak{T}) \neq 0, \Delta U(\mathfrak{T}) \neq 0$, allowing for side-effects, *'the work done on a substrate'* is the change $\delta W$ in the energy $U$ of the *work-media* required as side-effects in the construction; and the '*heat absorbed by the substrate*' is the change δQ in the energy *U* of the *heat media* required as side-effects. Whether or not such quantities are always positive (as under the known laws of physics), negative, or lack a definite sign, is for the subsidiary theory to decide. Constructor theory allows all such possibilities.

By focusing on side-effects being instantiated in heat media or work media – i.e. two different kinds of "agents of transfer" of energy, constructor theory is faithful to the traditional formulation, while also improving on it. For example, consider [36]:

"Energy has been transferred from source to object *through the agency of heat*: heat is the *agent of transfer*, not the entity transferred."

In constructor theory, the crucial difference is that such different 'agents of transfer' (work media and heat media) are distinguished from one another in a scale-independent way. This comes from the interoperability of work and heat – properties that cannot be stated in the traditional conception, but have elegant expressions in constructor theory. By the conservation of energy applied to the whole system, including the substrate in question *and* the side-effects of the construction task, one recovers the traditional formula, thus ensuring consistency with existing thermodynamics:

$$\Delta U = \delta W + \delta Q$$

In constructor theory, this formula is the culmination of the construction of thermodynamics, rather than its foundation; but it has now no circularity and its quantities are scale-independent, as promised.

## 8. Conclusions

This paper proposes three main results: a scale-independent, non-probabilistic formulation of the laws of thermodynamics in constructor theory, via the definition of *adiabatic possibility* in terms of possible computations; a scale-independent connection between the *first law of thermodynamics* and information theory; a scale-independent distinction between work and heat, rooted in the notion of distinguishability, which leads to the unification of Kelvin's statement of the second law and its axiomatic formulations. Constructor theory improves on the axiomatic approaches. The latter, too, impose restrictions on subsidiary theories via principles, or axioms, that aim to make contact with the physical world; however, those principles are scale-dependent, and rely on ad-hoc definitions, such as that of mechanical systems. In addition, their emphasis is on defining the entropy function of state. In constructor theory the principles are scale-independent, more general, and their physical content resides in the interoperability laws and in the notions of heat and work media, rather than in the existence of an entropy function.

This theory, being scale-independent, can be applied to single systems and to objects that are out of equilibrium (for which thermodynamic attributes are possible). Although beyond the scope of this paper, equilibrium thermodynamics can be recovered within this picture; and connections with the recently emerged field of quantum thermodynamics can be expected to arise. These are among the promising new avenues opened up by this approach.

**Acknowledgements**

I thank Raam Uzdin for several valuable comments; Harvey Brown for helpful discussions on axiomatic thermodynamics; Vlatko Vedral for several illuminating discussions about quantum thermodynamics; Douglas Moore and Sara Walker for valuable suggestions; and Peter Vadasz for many helpful comments about how to improve the presentation of this paper. I am grateful to Peter Atkins for deep discussions on classical thermodynamics; and for merciless, but constructive, criticism on this work. My special thanks and appreciation to David Deutsch, for numerous far-reaching discussions, and for providing inspiration and incisive criticism at every stage of this work. This publication was made possible through the support of a grant from the Templeton World Charity Foundation. The opinions expressed in this publication are those of the author and do not necessarily reflect the views of Templeton World Charity Foundation.

**Data accessibility**

This paper has no data.

**Competing Interest**
I have no competing interests.

**Authors' contributions**
The Author was the sole contributor to the content of this paper.

**Funding Statement**
The Author was supported by the Templeton World Charity Foundation.

**Ethics statement**
This work did not involve any collection of human data.

**References**

1. Deutsch, D. *"Constructor Theory"*, Synthese 190, 18, 2013.
2. Deutsch, D. Marletto, C. *"Constructor Theory of Information"*, Proc. R. Soc. A, 471:20140540, 2015.
3. Carathéodory, C., "Untersuchung über die Grundlagen der Thermodynamik" Mathematische Annalen, 67, 355–386 (1909).
4. Lieb, E., and Yngvason, "The Physics and Mathematics of the Second Law", J., Physics Reports, 310, 1–96 (1999), erratum, 314, 669, (1999).
5. Landauer, R. *'Irreversibility and heat generation in the computing process'*, IBM Journal of Research and Development, 183–191, (1961).
6. Bennett, C. H. (1987). *"Demons, engines and the second law"*. Scientific American, 257, 108–116.
7. Coecke, B. , Fritz, T.; Spekkens, R. *"A Mathematical theory of resources"*, arXiv:1409.5531v3 [quant-ph]
8. Abramsky, S., Coecke, B. *"Categorical quantum mechanics"* In: Kurt Engesser, Dov M.Gabbay & Daniel Lehmann, editors: *"Handbook of quantum logic and quantum structures: quantum logic"*, Elsevier, pp. 261–324, 2008.
9. Baylin, M. *"A survey of thermodynamics"*, New York 1994.
10. Buchdahl, H.A., *"The Concepts of Classical Thermodynamics,"* Cambridge University Press (1966).
11. Marsland, R.; Brown, H ., Valente G., Am. J. Phys., Vol. 83, No. 7, July 2015
12. Atkins, P. W. *"Physical Chemistry"*, Oxford University Press, 1998.

13. Uffink, J. *"Bluff your way to the Second Law of Thermodynamics"*, Studies in History and Philosophy of Science Part B 32 (3):305-394 (2001)

14. Goold, J., Huber, M., Riera, A., del Rio, L. & Skrzypczyk, P. *"The role of quantum information in thermodynamics—a topical review"*. J. Phys. A: Math. Theor. 49, 143001 (2016).

15. Brandão, F. G. S. L., Horodecki, M., Oppenheim, J. & Wehner, S. *"The second laws of quantum thermodynamics"*. Proc. Natl Acad. Sci. USA 112, 3275–3279 (2015).

16. Brown, H., Uffink, J., "The origin of time asymmetry in thermodynamics: the minus-first law", Studies in History and Philosophy of Science Part B 32 (4):525-538 (2001)

17. Feynman, R., Statistical Mechanics, Princeton University Press 1972.

18. Brown, H. R., Myrvold, W. and Uffink, J. (2009). *"Boltzmann's H-theorem, its discontents, and the birth of statistical mechanics"*. Studies in History and Philosophy of Modern Physics 40, 174‑191.

19. Linden, N., Popescu, S., Short, A. J. & Winter, A. *"Quantum mechanical evolution towards thermal equilibrium"*. Phys. Rev. E 79, 061103 (2009).

20. Landsberg, P. T., *"Thermodynamics and statistical mechanics"*, Dover Publications, 1978.

21. Wallace, D. *"The Quantitative Content of Statistical Mechanics"*, Studies in the History and Philosophy of Modern Physics 52 (2015) pp.285--293.

22. Wallace, D. *"Recurrence Theorems: a Unified Account"*, Journal of Mathematical Physics 65 (2015) 022105.

23. Leff, H. S., & Rex, A. F., (Eds.) (1990). *"Maxwell's demon: Entropy, information, computing"*. Adam Hilger, Bristol.

24. Earman, J., & Norton, J. D. (1999). Exorcist XIV: "The wrath of Maxwell's demon. Part II". Studies in History and Philosophy of Modern Physics, 30, 1–40.

25. Bekenstein, J.D., "Black holes and the second law", Lettere al Nuovo Cimento, 4, 737, (1972)

26. Atkins, P. W. *The Four Laws That Drive the Universe*, Oxford University Press, 2007.

27. Timpson, C., *"Quantum Information Theory and the Foundations of Quantum Mechanics"*, OUP, 2013.

28. Einstein A. 1970 *"Albert Einstein: philosopher, scientist", 3rd edn (ed. Schilpp PA),* p. 85. Evanston, IL: Library of Living Philosophers.
29. Deutsch D, Hayden P. 2000 "Information flow in entangled quantum systems". *Proc. R. Soc. Lond. A* **456**, 1759–1774
30. Barbour, J.; Koslowski,T. and Mercati, F., Phys. Rev. Lett. 113, 181101 – 2014.
31. Marletto, C.; Vedral, V. "Evolution without evolution and without ambiguities", arxiv: 1610.04773
32. Wootters W, Zurek W. 1982 "A single quantum cannot be cloned." *Nature* **299**, 802–803
33. Skrzypczyk, P., Silva, R. and Brunner N., "Passivity, complete passivity, and virtual temperatures ", Phys. Rev. E 91, 052133 (2015).
34. R. Giles, Mathematical Foundations of Thermodyamics, Pergamon Press, Oxford, 1964.
35. Marletto, C., "Constructor theory of Probability", Proc R Soc A **472**, 2192, 2016.
36. Atkins, P. W. *"Galileo's Finger"*, Oxford University Press, 2004.